\documentclass[prd,twocolumn,superscriptaddress]{revtex4-1}

\usepackage{graphicx, amsfonts, amsmath, amssymb, amscd,  theorem, exscale,dsfont,
epic, eepic, epsfig,enumerate}

\usepackage{hyperref} %

\usepackage{makeidx}
\makeindex

\newcounter{Figure}
\theoremstyle{plain}
\theoremheaderfont{\scshape}

\newtheorem{Def}{\bf Definition}

\newtheorem{Lem}{\bf Lemma}
\newtheorem{prop}{\bf Proposition}
\theorembodyfont{\upshape}

\theoremheaderfont{\scshape\small} \theorembodyfont{\scshape\small}

\newcommand{\qed}{\hspace{14cm} $\blacksquare$ \\} 

\newcommand{\nlap}{\mbox{$ \nabla \mkern -13mu / \ $}}

\newcommand{\be}{\begin{equation}}
\newcommand{\ee}{\end{equation}}
\newcommand{\bea}{\begin{eqnarray}}
\newcommand{\eea}{\end{eqnarray}}
\newcommand{\beas}{\begin{eqnarray*}}
\newcommand{\eeas}{\end{eqnarray*}}

\newcommand{\tr}{\text{tr} \,}
\newcommand{\curl}{\text{curl} \,}

\begin{document}

\title{Brill Waves with Slow Fall-Off Towards Spatial Infinity}

\author{Lydia Bieri}
\email{lbieri@umich.edu}
\affiliation{Department of Mathematics, University of Michigan, Ann Arbor, MI 48109-1120, USA}

\author{David Garfinkle}
\email{garfinkl@oakland.edu}
\affiliation{Dept. of Physics, Oakland University,
Rochester, MI 48309, USA \\ 
and Michigan Center for Theoretical Physics, Randall Laboratory of Physics, University of Michigan, Ann Arbor, MI 48109-1120, USA}

\author{James Wheeler}
\thanks{Corresponding author}
\email{jcwheel@umich.edu}
\affiliation{Department of Mathematics, University of Michigan, Ann Arbor, MI 48109-1120, USA}

%%%%%%%%%%%%%%%%%%%%%%%%%%%%%%%%%%%%%%%%%%%%%%%%%%%

\date{\today}

%%%%%%%%%%%%%%%%%%%%%%%%%%%%%%%%%%%%%%%%%%%%%%%%%%%

\begin{abstract} 
We compute families of solutions to the Einstein vacuum equations of the type of  Brill waves, but with slow fall-off towards spatial infinity. We prove existence and uniqueness of solutions for physical data and numerically construct some representative solutions. We numerically construct an explicit example with slow-off which does not exhibit antipodal symmetry at spatial infinity. 
\end{abstract}

\maketitle

\tableofcontents

\section{Introduction}
\label{intro}

Brill waves are simple solutions to the Einstein vacuum constraint equations. Nonetheless, they show interesting structures and allow one to study more involved properties of spacetimes in General Relativity. In this article, we compute families of solutions to the Einstein vacuum equations of the Brill wave type, but with slower fall-off towards spatial infinity than usually considered. 

Brill wave solutions were introduced by D. Brill in \cite{Brill1}.  See also \cite{omua} by N. \'O Murchadha, or \cite{khir} by A. Khirnov and T. Ledvinka for further results and discussions. They have been used as initial data in several numerical evolutions.  See {\it e.g.} \cite{garfdunc,alcubierreetal,rinne,hilditchetal}.

The Einstein vacuum equations are 
\be \label{EV}
R_{\mu \nu} = 0  \ 
\ee 
with $\mu, \nu = 0,1,2,3$. Generally, Greek indices denote spacetime coordinates, whereas Latin indices denote spatial components. 
We solve the equations for the unknown metric $\mathbf{g_{\mu \nu}}$, 
denoting the solution spacetimes by $(M, \mathbf{g})$. These are $4$-dimensional manifolds with a Lorentzian metric $\mathbf{g}$ solving the system of equations (\ref{EV}). We denote initial data for the Einstein equations by $(H, g_{ij}, k_{ij})$ with $i, j$ denoting spatial indices, where $H$ is a $3$-dimensional Riemannian manifold with metric $ g_{ij}$ and extrinsic curvature $k_{ij}$. Initial data have to satisfy the constraint equations 
(\ref{constr1})-(\ref{constr2}) below. 

Denote by $(T, e_1, e_2, e_3)$ a frame field for the spacetime $(M, \mathbf{g})$. Here $T$ is a future-directed unit timelike vector field, and the remaining vector fields form an orthonormal frame for the spacelike hypersurface $H$. 
In the following, $t$ and $\alpha$ are the time and lapse functions associated to the foliation of $M$ constructed by evolving the initial data. Curvature quantities with a bar refer to curvature in $H$, whereas $R_{\alpha \beta \gamma \delta}$ is the spacetime Riemannian curvature tensor. The same rules apply to the Ricci and scalar curvatures, as well as the differential operators.  

Let $k_{ij}$ denote the second fundamental form with respect to the $t$-foliation, given by 
\be \label{kglobaldef}
k_{ij} = - g (D_{e_i} T , e_j) \ . 
\ee
In particular, $k_{ij}$ obeys the first variation equation 
\be  \label{g1} 
\frac{\partial g_{ij}}{\partial t}  =  - 2 \alpha k_{ij}   \ . 
\ee

The structure equations ((\ref{k1})-(\ref{G2})) of the foliation read 
\bea
\frac{\partial k_{ij}}{\partial t} & = & - \bar{\nabla}_i \bar{\nabla}_j \alpha  \nonumber \\ 
& & \hspace{- 2cm}
+ ( R_{i0j0} - k_{im} k^m_{ \ j} ) \alpha   \label{k1}   \\ 
\bar{\nabla}_i k_{jm}  - \bar{\nabla}_j  k_{im} & = & R_{m0ij} \label{C1} \\ 
\bar{R}_{ij} + (\tr k) k_{ij}  - k_{im} k^m_{ \ j} & = & R_{ij} + R_{i0j0} \label{G2}  
\eea
The latter is the trace of the Gauss equation 
\be \label{G1}
\bar{R}_{imjn} + k_{ij} k_{mn} - k_{in} k_{mj}  =  R_{imjn} 
\ee

From these, one may derive the constraint and evolution equations for the Einstein vacuum (EV) equations (\ref{EV}): 

Constraint equations for EV
\bea
(\text{div} \, k)_j & = & \partial_j(\tr k ) \label{constr1} \\ 
\bar{R} + (\tr k)^2 - |k|^2 & = & 0 \label{constr2}  
\eea 

Evolution equations for EV 
\bea
\frac{\partial \bar{g}_{ij}}{\partial t} & =  & - 2 \alpha k_{ij}  \label{evol1}  \\ 
\frac{\partial k_{ij}}{\partial t} & = & 
- \bar{\nabla}_i \bar{\nabla}_j \alpha \label{evol2}
\\ & & \;\; + ( \bar{R}_{ij} + (\tr k) k_{ij}  - 2 k_{im} k^m_{ \ j}   ) \alpha \nonumber
\eea

We also obtain
\be \label{trkt1}
\frac{\partial \, \tr k}{\partial t} = - \Delta \alpha - ( \bar{R}  +(\tr k)^2  )  \alpha \ . 
\ee

Moreover, we have 
the following equations in $H$ 
\bea
(\curl k )_{ij} \ & = & \ H_{ij}   \label{k3}  \\ 
\bar{R}_{ij} \ & = & \ k_{im} k^m_{\ j} \ + \ E_{ij}  \ .  \label{k4} 
\eea
where $E_{ij}$ and $H_{ij}$ denote electric and magnetic parts of the Weyl curvature, respectively. 

When choosing, say, a maximal time function $t$, then $\tr k = 0$ and the above equations simplify accordingly. For the Brill-type solutions considered in the upcoming section, we will have $k = 0$, so the equations will simplify considerably, allowing 
simpler characterization of the corresponding initial data. 

The remainder of this article is organized as follows: In section \ref{Brill}, we briefly review the Brill waves as introduced in \cite{Brill1}, and give an overview of different types of asymptotically-flat spacetimes, from which data for the new Brill wave solutions are extracted. Existence and uniqueness of these new Brill wave solutions are then proven in section \ref{proof} and computed in section \ref{Brillnew}. The conclusions follow in section \ref{conclusions}. 

We note that as a byproduct we construct an example of Brill wave initial data whose evolution cannot satisfy the antipodal conjecture of \cite{strominger}. Recently, this topic has been discussed in the literature a lot, see for instance \cite{symmprabhuetal}, \cite{symmmagdyetal}, and a forthcoming paper addressing the general case \cite{lz1}.

\section{Brill Waves and General Relaxed Fall-Off Behavior}
\label{Brill}

In \cite{Brill1}, D. Brill studied a family of solutions of the Einstein equations that have since been called Brill waves. More specifically, Brill waves are solutions of the Einstein vacuum equations that are axisymmetric and time-symmetric. 
They are defined by the initial spatial metric on $H = \mathbb R^3$ given in cylindrical coordinates $\{ \rho, z, \varphi  \}$ by 
\be \label{Brillmetric1}
g = \Psi^4 ( e^{2q} (d \rho^2 + dz^2) + \rho^2 d \varphi^2  )  \ . 
\ee
The function $q = q(\rho, z)$ can be chosen, subject to mild conditions, and the conformal factor $\Psi$ is then determined by the constraint equations. In particular, we require 
\be
q = \frac{\partial q}{\partial \rho} = 0  \ \ \ \ \mbox{along the $z$-axis, }
\ee
that it decays fast enough at infinity, and that it is sufficiently regular. 

The time-symmetry implies that 
the extrinsic curvature vanishes, $k_{ij} = 0$. 
As a result, the constraint equation (\ref{constr1}) becomes trivial, whereas (\ref{constr2}) simplifies to 
\be \label{constr23}
\bar{R} = 0 \ . 
\ee
Due to time symmetry, then, the momentum constraints are trivial, so to construct the data we only need to solve the Hamiltonian constraint (\ref{constr23}). Under the metric ansatz (\ref{Brillmetric1}), this reduces to
\be \label{reducedPsi}
\Delta \Psi + \frac{1}{4} \left( \frac{\partial^2 q}{\partial \rho^2} + \frac{\partial^2 q}{\partial z^2} \right) \Psi  = 0 \ . 
\ee

In \cite{Brill1}, Brill looked at data that behaves towards spatial infinity like 
\bea
q & = & O (r^{-2} ) \label{brillfalloff1}  \\ 
\Psi & = & 1 + \frac{M}{2r} + O ( r^{-2} ) \label{brillfalloff2}\ 
\eea
with the constant $M$ being the total mass. Brill's motivations for considering such data included both that it was broadly considered a reasonable starting point in early analyses of spacetime asymptotics and, more pointedly toward his paper's objective, that it lent itself to a simple and explicit characterization of $M$.

It is now known that the asymptotics (\ref{brillfalloff1})-(\ref{brillfalloff2}) are more stringent than necessary to define physical notions such as mass and angular momentum and enjoy stability properties (\cite{lydia1,lydia2,lydia4,lydia14,sta}). In an effort to understand the features of broader physically admissible spacetimes, then: in this article, we compute Brill-type solutions for more general data with the slower fall-off (\ref{qA})-(\ref{PA}) and (\ref{qB})-(\ref{PB}), discussed below. Before turning to the Brill wave case, where $k_{ij} = 0$, we first discuss the more general data $k_{ij} \neq 0$ in a larger framework to set up the context.

Ultimately, we need to understand the behavior of our spacetimes towards infinity, in particular of the quantities $q$ and $\Psi$ above. 
As we study asymptotically-flat initial data in the following subsection, we shall briefly introduce different types of such data, collecting the fall-off behavior of both the data and the resulting spacetime curvature. These will yield the physical scenarios from which we shall extract new fall-off behavior for $q$ and $\Psi$ in the Brill-wave ansatz. 

\subsection{Asymptotically-Flat Spacetimes}
\label{datatypes}

{\itshape Data of type (A):} 
First, we consider initial data $(H, g_{ij}, k_{ij})$ 
for which there exists a coordinate system $(x^1, x^2, x^3)$ in a neighborhood of infinity in which the metric and extrinsic curvature satisfy,  as 
$r := (\sum_{i=1}^{3} (x^i)^2 )^{\frac{1}{2}} \to \infty$,  
\bea
g_{ij} \ & = & \ \delta_{ij} \ + \ h_{ij} \  + \ o_3 \ (r^{- \frac{3}{2}})  \label{initialdg1}  \\ 
k_{ij} \ & = & \ o_2 (r^{-\frac{5}{2}})  \label{initialdk1}   
\eea
with $h_{ij}$ being homogeneous of degree $-1$. 
In particular, $h$ may include a non-isotropic mass term $M(\theta, \phi)$ depending on the angles $\theta,\phi$. 
The spacetime metric will include a resulting term, being homogeneous of degree $-1$ with corresponding limit $M(u, \theta, \phi)$ at future null infinity depending on the retarded time $u$.  We refer to this type of initial data and the corresponding {spacetimes} as {(A)}.

{\scshape Notation:} By $u$ we denote the optical function (as in the stability proofs \cite{sta}, \cite{lydia1}, \cite{lydia2}) corresponding to {\itshape minus the retarded time} in Minkowski spacetime. We refer to $u$ just as the {\itshape retarded time} with this sign convention. Set $\tau_- := \sqrt{1 + u^2}$. Furthermore, we say that $f = o_m(r^\delta)$ provided that $D^\alpha f = o(r^{\delta-|\alpha|})$ for any multi-index $\alpha$ of order $|\alpha| \leq m$

At this point, we refer the reader to appendix \ref{appinfo} for the definitions of the curvature components to be discussed. It is shown in \cite{lydia4,lydia14,lydia5} that, for {spacetimes of type (A)},
the components of the spacetime curvature behave as 
\bea
\underline{\alpha} \ & = & \ O \ ( r^{- 1} \ \tau_-^{- \frac{5}{2}}) \\ 
\underline{\beta} \ & = & \ O \ ( r^{- 2} \ \tau_-^{- \frac{3}{2}}) \\ 
\varrho 	\ & = & \ O \ ( r^{- 3})  \label{rhoA1} \\ 
\varrho - \bar{\varrho} \ & = & \ O \ ( r^{- 3})  \label{rhoA2} \\ 
\sigma \ & = & \ O \ ( r^{- 3} \ \tau_-^{- \frac{1}{2}}) \\ 
\sigma - \bar{\sigma}  \ & = & \ O \ ( r^{- 3} \ \tau_-^{- \frac{1}{2}}) \\ 
\beta \ & = & \ o \ (r^{- \frac{7}{2}})   \label{betaA}  \\ 
\alpha  \ & = & \ o \ (r^{- \frac{7}{2}})  
\eea

Note that in the initial hypersurface $H_0$, the decay behavior translates into $r^{- \frac{7}{2}}$ fall-off for all curvature components except for 
$ \varrho \  = \ O \ ( r^{- 3})  \label{rhoA1}$ and 
$\varrho - \bar{\varrho} \ = \ O \ ( r^{- 3})  \label{rhoA2}$. 
The analogous statement is true for the other spacetimes below, but with different fall-off.

In particular, we also derive for type (A) spacetimes that 
\bea
\nlap \varrho 	\ & = & \ O \ ( r^{- 4}) \label{nablarhoLB1}
\eea
whereas in the stability result by Christodoulou and Klainerman (CK) it is 
\bea
\nlap \varrho 	\ & = & \ O \ ( r^{- 4} \tau_-^{- \frac{1}{2}})  \label{nablarhoCK}  \ . 
\eea

From this we extract the following fall-off conditions for the initial data to Brill-wave solutions of type (A), which we wish to investigate: 
as $r \to \infty$,
\bea
q & = & O (r^{-3/2} )  \label{qA} \\ 
\Psi & = & 1 + \frac{M}{2r} + O (r^{-3/2} ) \ .  \label{PA}
\eea
Here, $M$ is a mass term which may, in general, be non-isotropic, having  $M(\theta, \phi)$ depend on angle. In section \ref{proof}, however, 
we prove that any solutions of this type must, in fact, have $M$ constant, reducing the data to the following, type (CK).

{\itshape Data of type (CK):} 
In \cite{sta}, D. Christodoulou and S. Klainerman considered (CK) data, refer to this type of initial data and the corresponding {spacetimes} as {(CK)}: 
\bea
g_{ij} \ & = & \ \left(1 \ + \ \frac{2M}{r} \right) \ \delta_{ij} \ + \ o_4 \ (r^{- \frac{3}{2}}) \label{safg33} \\ 
k_{ij} \ & = & \  o_3 \ (r^{- \frac{5}{2}}) \ ,  \label{safk33}
\eea
where $M$ denotes the mass and is constant.

For {spacetimes of type (CK)} the curvature components behave like 
\bea
\underline{\alpha} \ & = & \ O \ ( r^{- 1} \ \tau_-^{- \frac{5}{2}}) \\ 
\underline{\beta} \ & = & \ O \ ( r^{- 2} \ \tau_-^{- \frac{3}{2}}) \\ 
\varrho 	\ & = & \ O \ ( r^{- 3})  \label{rhoCK1} \\ 
\varrho - \bar{\varrho} \ & = & \ O \ ( r^{- 3} \ \tau_-^{- \frac{1}{2}}) \label{rhoCK2} \\ 
\sigma \ & = & \ O \ ( r^{- 3} \ \tau_-^{- \frac{1}{2}}) \\ 
\sigma - \bar{\sigma}  \ & = & \ O \ ( r^{- 3} \ \tau_-^{- \frac{1}{2}}) \\ 
\beta \ & = & \ o \ (r^{- \frac{7}{2}}) \label{betaCK} \\ 
\alpha \ & = & \ o \ (r^{- \frac{7}{2}})
\eea

{\itshape Data of type (B):} 
In \cite{lydia1}, \cite{lydia2}, L. Bieri proved stability for initial data of the following type, 
referring to this type of initial data and the corresponding {spacetimes} as {(B)}: as $r \to \infty$,
\bea
g_{ij} \ & = & \ \delta_{ij} \ + \ 
o_3 \ (r^{- \frac{1}{2}}) \label{LBg}  \\
k_{ij} \ & = & \ o_2 \ (r^{- \frac{3}{2}})   \ .      \label{LBk}  
\eea

Further, in the initial hypersurface $H_{0}$ it follows that 
the Weyl curvature $W$ is of the order $r^{- \frac{5}{2}}$.  
At this point, recall equation (\ref{k4}) for the spatial Ricci curvature. It then follows from the above as well as  (\ref{k4}), (\ref{LBk}) that 
the spatial Ricci curvature $\overline{Ric}$ is of the order  $r^{- \frac{5}{2}}$. 
From (\ref{LBk}) and (\ref{constr2}) with $\tr k=0$ it follows that 
\be 
\bar{R}  \ =  \  |k|^2 \ = \ o(r^{-3})   \label{intRbarLB1} \ \ . 
\ee
Thus, summarizing, we have in $H_{0}$: 
\bea
W & = & o(r^{- \frac{5}{2}}) \\ 
\overline{Ric} & = & o(r^{- \frac{5}{2}}) \\ 
\bar{R} & = &  o(r^{-3}) \ \ . 
\eea

It was shown in \cite{lydia1}, \cite{lydia2} that in spacetimes $(M,\mathbf{g})$ of type (B), the curvature components behave as follows: 
\bea
\underline{\alpha} \ & = & \ O \ ( r^{- 1} \ \tau_-^{- \frac{3}{2}}) \\ 
\underline{\beta} \ & = & \ O \ ( r^{- 2} \ \tau_-^{- \frac{1}{2}}) \\ 
\varrho , \ \sigma , \ \alpha , \ \beta \ & = & \ o \ (r^{- \frac{5}{2}})  \ . 
\eea

From this we extract the following fall-off conditions for the initial data to Brill-wave solutions of type (B), which we wish to investigate: 
as $r \to \infty$,
\bea
q & = & O(r^{-1/2})   \label{qB}  \\ 
\Psi & = & 1  + O(r^{-1/2}) \ .   \label{PB}
\eea
In section \ref{Brillnew}, we present computations for a concrete example of this type.

\section{Existence and Uniqueness of Brill Wave Solutions with Relaxed Fall-Off Behavior}
\label{proof}
We would like to compute and characterize new Brill wave solutions both of type (A) and of type (B), and this amounts to constructing a pair $q, \Psi$ satisfying (\ref{qA})-(\ref{PA}) or (\ref{qB})-(\ref{PB}), respectively. For this purpose, we return to the situation where $k_{ij} = 0$ as explained in the introduction, and recall that the constraint equations reduce to (\ref{constr23}), giving (\ref{reducedPsi}). Setting 
\be
\phi := -\frac{1}{4} \left( \frac{\partial^2 q}{\partial \rho^2} + \frac{\partial^2 q}{\partial z^2} \right),
\label{phidef}
\ee
we see that (\ref{reducedPsi}) is equivalent to the Schr\"odinger-type boundary value problem (BVP)
\bea \label{schrodingerPsi}
-\Delta \Psi + \phi \Psi & = & 0     \notag
\\ \lim_{r \to \infty} \Psi & = & 1 \ ,
\eea
with ``potential" $\phi$. As in \cite{Brill1}, our strategy is to specify an admissible $q$ and construct a corresponding $\Psi$ by solving (\ref{schrodingerPsi}). It is not obvious a priori precisely when this can be done, or that the resulting $\Psi$ will have the desired structure and decay. In this subsection, we establish that it can be done in principle, and done uniquely, under reasonable conditions on $q$, and that the constructed $\Psi$ will satisfy appropriate decay conditions. Namely: up to a mass term, $\Psi-1$ decays at least as quickly as $q$, so that the desired decay rate can be easily prescribed.

To bring analytical tools to bear, we operate in the weighted Sobolev spaces $W^{k,p}_\delta$ and $W^{\prime \, k,p}_\delta$ ($k \in \mathbb N_0$, $p \in [1,\infty)$, and $\delta \in \mathbb R$) on $\mathbb R^n$, function spaces defined and reviewed in Appendix \ref{app:sobolev}. Roughly, functions in $W^{k,p}_\delta$ or $W^{\prime \, k,p}_\delta$ should be thought of as decaying at least as quickly as $r^\delta$ as $r \to \infty$, with each derivative up to order $k$ having an additional power of fall-off (the difference between the primed and unprimed spaces is a minor technical point involving integrability around the origin). Indeed: if $kp > n$, then $U \in W^{k,p}_\delta \implies U = o(r^\delta)$ (Proposition \ref{prop:weighted}). 

Although perhaps unwieldy relative to the more familiar unweighted Sobolev spaces $W^{k,p}$, the weighted spaces are essential to the present analysis, its core objective being to track and control fall-off behavior. Knowing that quantities involving $\Psi$ belong to $W^{k,p}_\delta$ for differing $\delta$, for example, can indicate whether one has (\ref{PA}), (\ref{PB}), or something in between. Our first objective is to show that the fall-off of $\Psi-1$ must appropriately follow that of $q$, drawing a correlation between weighted spaces containing $\Psi-1$ (or related quantity) and those containing $q$.

Toward this end: given $q \in W^{2,p}_{-\tau}$ for some $p > n/2$ and $\tau > 0$, we clearly have $\phi \in W^{0,p}_{-2-\tau}$, and it is convenient to identify the linear differential operator 
\be
P := \Delta - \phi,
\ee
which may be considered as a bounded linear operator $W^{2,p}_\delta \to W^{0,p}_{\delta-2}$ for any $\delta \in \mathbb R$. $P$ is clearly asymptotic to $\Delta$ at rate $\tau$ (Definition \ref{def:asymptotic}). To naturally incorporate the boundary condition $\Psi \to 1$, it is further convenient to set $U := \Psi - 1$ and recast (\ref{schrodingerPsi}) as
\bea \label{schrodingeru}
  PU & = & \phi.    \notag
\\ U & = & o(1).    
\eea
This is the form we shall analyze. While the present work is primarily concerned with applications to $n = 3$, results in this section will be general to $n \geq 3$.

The following result establishes that any solutions must decay appropriately.
\begin{prop} \label{prop:decay}
Suppose that a linear differential operator $L$ is asymptotic to $\Delta$ at rate $\tau>0$ and exponent $2p > n$, $f \in W^{0,p}_{\delta'-2}$, and that $U\in W^{2,p}_{\delta}$ ($\delta' \leq \delta \in \mathbb R$ both non-exceptional) solves $LU = f.$ Then there is an exceptional $k \leq k^-(\delta)$, a harmonic function $h_k$ of order $r^{k}$, and an $\eta \in C^1(\mathbb R^n \backslash \{0\})$ satisfying $\eta(x) = 0$ for $|x| > 1$ such that
\beas
U - h_k - \eta & \in & \left( \bigcap_{\epsilon > 0} W^{2,p}_{k+\epsilon-\tau} \right) \cup W^{2,p}_{\delta'} \\ & = & \bigcap_{\epsilon > 0} W^{2,p}_{\max(k+\epsilon-\tau,\delta')}.
\eeas
In particular, 
\be
U - h_k = o(r^{\max(k+\epsilon-\tau, \delta')})
\ee
for every $\epsilon > 0$.
\end{prop}
Here, $\delta \in \mathbb R$ is called {\it exceptional} if it is among the possible growth rates of harmonic functions in dimension $n$, i.e.\@ if $\delta \in \mathbb Z \backslash \{ 2-n < k < 0\}$, and 
$$k^-(\delta) := \max \{ k < \delta \, \big | \, k \text{ exceptional}\}.$$ 

This proposition, the explicit statement of an inhomogeneous version of Theorem 1.17 of \cite{bartnik1} (see also \cite{meyers}), essentially states that the deviation of a solution $U$ from a harmonic function is controlled at infinity by either the operator $L$'s deviation from $\Delta$ or the fall-off of the source term $f$, whichever is weaker. The proof requires the following simple existence lemma for the Laplacian on $W^{k+2,p}_\delta$, drawing on the fact that $\Delta: W^{\prime \, k+2,p} _{\delta} \to W^{\prime \, k,p} _{\delta-2}$ is an isomorphism for $\delta$ non-exceptional and $1 < p < \infty$ (\cite{bartnik1}, Theorem 1.7).

\begin{Lem} \label{lem:surjective}
Given $R > 0$ and $F \in W^{k,p}_{\delta-2}$ with $k \in \mathbb N_0$, $1 < p < \infty$ and $\delta$ non-exceptional, there exists a $U \in W^{k+2,p}_\delta$ such that
$$ \Delta U = F \quad \text{on} \quad |x|>R.$$
\end{Lem}
{\it Proof.} Let $\chi \in C^\infty(\mathbb R^n)$ satisfy $\chi(x) = 1$ on $\mathbb R^n \backslash B_1$ and $\chi(x) = 0$ on $B_{1/2}$, and set $\chi_R(x) := \chi(x/R)$. Then $\chi_R F \in W^{\prime \, k,p} _{\delta-2}$, so there is a $U' \in W^{\prime \, k+2,p} _{\delta}$ satisfying $\Delta U' = \chi_R F$. Take $U := \chi_R  U' \in W^{k+2,p}_\delta$. \qed

\noindent {\it Proof of Proposition \ref{prop:decay}}. Given a solution $U \in W^{2,p}_{\delta}$, we have 
\bea
\Delta U & = & LU + (\Delta - L) U \notag
\\ & = & f + (\Delta - L) U =: F.
\eea
Set $\delta_m := \max(\delta-m \tau,\delta') \leq \delta$ for $m \in \mathbb N_0$, and assume (without loss of generality) these are all non-exceptional. As $\delta_0 = \delta$, we are given that $U \in W^{2,p}_{\delta_0}$. From the decay assumptions on $L - \Delta$ and $f$, we deduce $F \in W^{0,p}_{\delta_1-2}$. Lemma \ref{lem:surjective} ensures that there is a $v_1 \in W^{2,p}_{\delta_1}$ such that
\be
\Delta (U-v_1) = 0 \quad \text{on} \quad |x|>1,
\ee
so that
\be
U - v_1 = h_{k_0} \quad \text{on} \quad |x|>1
\ee
for some harmonic function $h_{k_0}$ of degree $k_0 \leq k^-(\delta_0) = k^-(\delta)$ by the decay properties of $U$ and $v$. If $\delta_1 < k_0$, then the harmonic term dominates at infinity and $U \in W^{2,p}_{k_0+\epsilon}$ (for any $\epsilon > 0$), and if instead $\delta_1 > k_0$, then $U \in W^{2,p}_{\delta_1}$. In the latter case, our hypothesis has reset, and we may repeat the above argument to find a $v_2 \in W^{2,p}_{\delta_2}$ such that 
\be
U - v_2 = h_{k_1} \quad \text{on} \quad |x|>1
\ee
with $k_1 \leq k^-(\delta_1) \leq k^-(\delta)$, similarly obtaining either $u \in W^{2,p}_{k_1+\epsilon}$ or $U \in W^{2,p}_{\delta_2}$. We iterate in this way $M$ times until either $U \in W^{2,p}_{k_M+\epsilon}$ for some $\delta_{M+1} < k_M < \delta_M$, or $\delta_M = \delta'$, so that $v_M \in W^{2,p}_{\delta'}$. The result is achieved in the latter case; in the former, we set $\delta_* := \max(k_M+\epsilon-\tau,\delta')$ and proceed one last time to find $F \in W^{2,p}_{\delta_*-2}$, and hence there is a $v_* \in W^{2,p}_{\delta_*}$ with
\be
U - v_* = h_{k_M} \quad \text{on} \quad |x|>1,
\ee
as desired. Note that the final harmonic term must be of degree $k_M$ (not less than $k_M$), or else it would contradict the known decay rate of $U$ from the $M$th iteration concluding $U = h_{k_M} + v_{M+1}$ on $|x| > 1$, in particular that $\lim_{r \to \infty} r^{-k_M} |U| \neq 0$.

Regularity of the error term $\eta$ follows from the usual Sobolev Embedding Theorem.
\\ \qed

Applying Proposition \ref{prop:decay} to our context, i.e.\@ to (\ref{schrodingeru}) with $P = \Delta - \phi$ in $n = 3$, we have $0< \delta < 1$ and $\delta' = -\tau$. The result initially yields $k \leq 0$ and $U - h_0 - \eta \in W^{2,p}_{\epsilon-\tau}$, but since $U = o(1)$, the degree 0 term in $h_0$ must be nill, improving the estimate to $U - h_{-1} - \eta \in W^{2,p}_{-\tau}$. In particular, for slow decay $0 < \tau < 2$,
\be
U = \frac{C}{r} + o(r^{-\tau}),
\ee
indicating that, up to a mass term, the fall-off of $U$ is controlled by that of $q$, as desired. Note that $C$ here must be constant, so Brill wave solutions of type (A) cannot have anisotropic mass.

It remains to establish when a unique solution for $U$ exists. Though a more general version of the following result was proven by Choquet-Bruhat and Christodoulou in \cite{choquet} (Theorem 6.6), we include a proof demonstrating that Proposition \ref{prop:decay} yields precisely the needed decay to carry through the argument simply in our context.

\begin{prop} \label{prop:isomorphism}
Given $\phi \in W^{0,p}_{-2-\tau}$ with $\tau >0$, $p > n/2$, and $\phi(x) \geq 0$, the linear differential operator $P: W^{2,p}_\delta \to W^{0,p}_{\delta-2}$ given by $P := \Delta - \phi$ is an isomorphism for $2-n < \delta < 0$.
\end{prop}

\noindent {\it Proof.} As $P$ is clearly self-adjoint, discussion surrounding Proposition 1.14 in \cite{bartnik1} indicates that $P$ is Fredholm of index $0$ for $2-n < \delta < 0$, hence surjectivity is equivalent to injectivity. We establish the latter.

Suppose $U \in W^{2,p}_\delta$ satisfies $PU = 0$, so that $\Delta U - \phi U = 0$. Take $(U_m)_{m=1}^\infty \subset C^\infty_c(\mathbb R^n)$ such that $U_m \to U$ in $W^{2,p}_\delta$, and observe
\bea
& 0 & = \int_{\mathbb R^n} (- \Delta U + \phi U ) U_m d^n x \notag
\\ & & = \int_{\mathbb R^n} (\nabla U \cdot \nabla U_m + \phi U U_m) d^n x \notag
\\ & & = \int_{\mathbb R^n} (|\nabla U|^2 + \phi U^2) d^n x
\\ & + & \int_{\mathbb R^n} \left[ \nabla U \cdot \nabla(U_m - U) + \phi U (U_m-U) \right ] d^n x. \notag
\eea
By Proposition \ref{prop:decay} with $\delta' \to -\infty$, there is an exceptional $k \leq k^-(\delta) = 2-n < 0$ such that 
\be 
U = O(r^{k}) , \quad |\nabla U| = O(r^{k-1}),
\ee
so $U \in W^{1,s}_{k+\epsilon}$ for any $\epsilon > 0$ and $s \in [1,\infty)$. This allows us to estimate the error terms above:
\bea 
& & \left | \int_{\mathbb R^n} \nabla U \cdot \nabla(U_m - U) d^n x\right |   \notag
\\ & \leq & \|\nabla U \cdot \nabla(U_m - U)\|_{1,-n} \notag
\\ & \leq & \|\nabla U\|_{p',1-n-\delta} \cdot \| \nabla(U_m-U) \|_{p,\delta-1}  \notag
\\ & \leq & \|U\|_{1,p',k-\delta} \cdot \| U_m-U \|_{2,p,\delta},
\eea
by the weighted H\"older's inequality (See Proposition \ref{prop:weighted} in Appendix \ref{app:sobolev}) and since $k \leq 2-n$. Here, $p'$ is defined by $\frac{1}{p'} = 1 - \frac{1}{p}$. Similarly,
\bea
& & \left | \int_{\mathbb R^n} \phi U (U_m-U) d^n x \right | \notag
\\ & \leq & \| \phi U (U_m-U) \|_{1,-n} 
\\ & \leq & \| \phi \|_{p,-2-\tau} \cdot \|U\|_{p'',2-n+\tau-\delta} \cdot  \| U_m-U \|_{p,\delta} \notag
\\ & \leq & \| \phi \|_{p,-2-\tau} \cdot \|U\|_{p'',k+\tau-\delta} \cdot  \| U_m-U \|_{p,\delta}, \notag
\eea
where $\frac{1}{p''} = 1 - \frac{2}{p}$. As $m \to \infty$, then, each of these tends to $0$, giving
\be
0 = \int_{\mathbb R^n} (|\nabla U|^2 + \phi U^2),
\ee
hence $\nabla U = 0$ a.e.\@ since $\phi \geq 0$, forcing $U \equiv 0$ and showing that $\ker(P)$ is trivial.
\\ \qed

This result implies that, under the stated hypotheses, $PU = \phi$ has a unique solution in $W^{2,p}_\delta$ for $\max(2-n,-\tau) < \delta < 0$, and Proposition \ref{prop:decay} ensures that this uniqueness is sufficient given $U = o(1)$. Note that in general, however, Proposition \ref{prop:isomorphism}'s claim may be false without the sign condition on $\phi$. One may construct a counterexample in $n = 3$ by taking, say, 
\be 
U(x) = \frac{1}{\sqrt{1+ r^2}},
\ee
which is strictly positive, everywhere smooth, and satisfies $U(x) = \frac{1}{r} + O(r^{-3})$, and defining 
\be
\phi(x) := \frac{\Delta U}{U} = \frac{-3}{(1+r^2)^2},
\ee
which is also everywhere smooth. Clearly $\phi \in W^{0,p}_{-2-\tau}$ for $0 < \tau < 2$, but $U \in \ker(\Delta - \phi)$ on $W^{2,p}_{\delta}$ for $-1 < \delta < 0$, so $\Delta-\phi$ is not an isomorphism. This $\phi(x)$, of course, violates the sign condition.

Regardless of sign, however, such counterexamples do not arise if $\phi$ (and hence $q$) is taken sufficiently ``small":

\begin{prop} \label{prop:isomorphismsmall}
Given $p > n/2$ and $2-n < \delta < 0$, there exists a constant $C > 0$ such that if $\phi \in W^{0,p}_{-2-\tau}$ with $\tau >0$ satisfies $\|\phi\|_{p,-2} < C$, the linear differential operator $P: W^{2,p}_\delta \to W^{0,p}_{\delta-2}$ given by $P := \Delta - \phi$ is an isomorphism.
\end{prop}

\noindent {\it Proof.} By Proposition \ref{prop:isomorphism} with $\phi \equiv 0$, $\Delta : W^{2,p} _{\delta} \to W^{0,p} _{\delta-2}$ is an isomorphism for these $\delta$, and Theorem 1.7 in \cite{bartnik1} demonstrated the boundedness of $K: W^{\prime \, 0,p} _{\delta-2} \to W^{\prime \, 2,p} _{\delta}$ given by
\be
(K f)(x) := - c_n \int_{\mathbb R^n} \frac{f(x')}{|x-x'|^{n-2}} d^n x'
\ee
with $c_n > 0$. Since $\delta - 2 < -2 < -n/p,$ a trivial comparison of norms yields $ W^{0,p} _{\delta-2} \subset W^{\prime \, 0,p} _{\delta-2}$ (note that combining with the weighted Poincar\'e inequality, Theorem 1.3 of \cite{bartnik1}, further gives $ W^{2,p} _{\delta} \subset W^{\prime \, 2,p} _{\delta}$), so we may restrict $K$ to a bounded operator $K: W^{0,p} _{\delta-2} \to W^{\prime \, 2,p} _{\delta}$. The operator
\be
K \circ \Delta :  W^{2,p} _{\delta} \to W^{\prime \, 2,p} _{\delta}
\ee
is then bounded, hence continuous, and it is a standard result that this restricts to the identity on $C^\infty_c(\mathbb R^n)$. We have by continuity, then, that $(K \circ \Delta)U = U$ for all $U \in W^{2,p} _{\delta}$, so the codomain of $K$ may be tightened to obtain
\be
K = \Delta^{-1}: W^{0,p} _{\delta-2} \to W^{2,p} _{\delta}
\ee
is a bounded isomorphism by the Inverse Operator Theorem.

Now, given $\phi \in W^{0,p}_{-2-\tau}$, we consider the linear differential operator $P := \Delta - \phi : W^{2,p}_\delta \to W^{0,p}_{\delta-2}$. Invertibility of $P$ is equivalent to that of 
\be
K \circ P = \mathds 1 - K_\phi : W^{2,p}_\delta \to W^{2,p}_\delta,
\ee
where $K_\phi : W^{2,p}_\delta \to W^{2,p} _{\delta}$ is defined by
\bea
& & (K_\phi U)(x) := (K (\phi U))(x)   \notag
\\ & & = -c_n \int_{\mathbb R^n} \frac{\phi(x') U(x')}{|x-x'|^{n-2}} d^n x'.
\eea
$\mathds 1 - K_\phi$ is now invertible, with inverse given by
\be
(\mathds 1 - K_\phi)^{-1} = \sum_{k=0}^\infty (K_\phi)^k,
\ee
provided that this series converges in $\mathcal L (W^{2,p}_\delta)$, in particular if $\|K_\phi\|_\text{op} < 1$ with $\|\cdot \|_\text{op}$ the operator norm. Utilizing the weighted H\"older's and Sobolev inequalities (Proposition \ref{prop:weighted}), the estimate
\bea
\| K_\phi u \|_{2,p,\delta} & \leq & \|K\|_\text{op} \|\phi u\|_{p,\delta-2}  \notag
\\ & \leq & \|K\|_\text{op} \|\phi\|_{p,-2} \|u\|_{\infty, \delta} \notag
\\ & \leq & \tilde C \|K\|_\text{op} \|\phi\|_{p,-2} \|u\|_{2,p, \delta}
\eea
(for some $\tilde C > 0$) yields 
\be
\|K_\phi\|_\text{op} \leq \tilde C \|K\|_\text{op} \|\phi\|_{p,-2},
\ee
so we may take $C = (\tilde C \|K\|_\text{op})^{-1}$. 
\\ \qed

Given $\phi \in W^{0,p}_{-2-\tau}$ satisfying this smallness condition, then, the problem $P U = \phi$ admits a unique solution $u \in W^{2,p}_\delta$ for $\max(2-n,-\tau) < \delta < 0$, given by
\be \label{seriessolution}
U = (\mathds 1 - K_\phi)^{-1}(K \phi) = \sum_{k=0}^\infty (K_\phi)^{k+1}(1).
\ee
From this we deduce
\be
|U| \leq \sum_{k=0}^\infty (K_{-|\phi|})^{k+1}(1),
\ee
and since this RHS increases with $|\phi|$, we see that we can ensure $|U| < 1$, so that $\Psi = 1+U > 0$ is an admissible conformal factor, if $-|\phi|$ can be bounded below by a $\phi_m$ for which the corresponding series solution $U_m$ in (\ref{seriessolution}) converges pointwise and satisfies $|U_m| < 1$. Minorly adjusting an argument of Brill \cite{Brill1}, this can be ensured if $\phi_m = \phi_m(r) \leq 0$ depends only on $r := |x|$, is continuous, and satisfies
\be
\int_0^\infty r |\phi_m(r)|dr < \frac{1}{2}.
\ee
Explicitly, one might take
\be
\phi_m(r) =  \begin{cases} 
      -M^2 & r \leq R \\
      -M^2 \left( \frac{R}{r} \right)^{2+\tau} & r \geq R
   \end{cases}
\ee
with $MR < \sqrt{\frac{\tau}{2+\tau}}$.

Taken together, Propositions \ref{prop:decay}-\ref{prop:isomorphismsmall} establish that the BVP (\ref{schrodingeru}) admits a unique solution for a broad class of admissible $q$, including an open ball around the origin of $W^{2,p}_{-\tau}$, and that any solution satisfies that the decay of $U$ is controlled by that of $q$. We are thereby assured that the task of numerically integrating (\ref{schrodingeru}) with appropriate, e.g.\@ sufficiently small, $q \in W^{2,p}_{-1/2}$ or $W^{2,p}_{-3/2}$ both is well-posed and will indeed yield initial data sets of type (A) or (B), respectively.

\section{Numerical Computations of Brill Waves with Relaxed Fall-Off Behavior} 
\label{Brillnew}

Finally, we are going to compute new Brill-wave solutions for the general data of types (A) as given in (\ref{qA})-(\ref{PA}) and 
for types (B) as given in (\ref{qB})-(\ref{PB}). We recall that, under time-symmetry, the constraint equations reduce to (\ref{constr23}) and we have (\ref{reducedPsi}), and ultimately the BVP (\ref{schrodingerPsi}).  
We follow the scheme of Section \ref{proof}: we pick the function $q$, use it to compute $\phi$ given by eqn.\@ (\ref{phidef}) and then numerically solve eqn.\@ (\ref{schrodingerPsi}).  We will find it convenient to perform the numerical computations in spherical polar coordinates $(r,\theta)$ which are related to the cylindrical coordinates by $z=r\cos \theta, \; \rho=r \sin \theta$.  In that case, the formula for $\phi$ given in eqn. (\ref{phidef}) becomes

\be
\phi = - {\frac 1 4} \left ( {\frac {{\partial ^2}q} {\partial {r^2}}} + {\frac 1 r} \, {\frac {\partial q} {\partial r}} + {\frac 1 {r^2}} \, {\frac {{\partial ^2}q} {\partial {\theta ^2}}} \right ) 
\ee

We define $F \equiv \ln \Psi$ and define $F_1$ and $F_2$ by $F={F_1}+{F_2}$ and 
\be
\Delta {F_1} = \phi
\label{F1eqn}
\ee
Then eqn. (\ref{schrodingerPsi}) becomes
\be
\Delta {F_2} = - {\vec \nabla}F \cdot {\vec \nabla}F
\label{F2eqn}
\ee
Thus given a numerical method to invert the Laplacian we first solve eqn. (\ref{F1eqn}) and then solve eqn. (\ref{F2eqn}) by iteration.  That is, having found $F_1$, we make the initial guess of zero for $F_2$ and then repeatedly solve eqn. (\ref{F2eqn}) for an improved version of $F_2$, where the right hand side of eqn. (\ref{F2eqn}) is computed using the previous version of $F_2$.  At each step, the current version of $F_2$ is stored, with the final version used to compute $F$.

Our numerical method to invert the Laplacian is the standard Green's function for axisymmetric functions (see, e.g. \cite{Jackson}) where we compute all integrals numerically. Explicitly, for eqn. (\ref{F1eqn}) we have 
\be
{F_1} = {\sum _{\ell =0} ^\infty} \left [ {g_\ell}(r) {r^{-(\ell+1)}} + {h_\ell}(r) {r^\ell}\right ] {P_\ell}(\cos \theta )
\ee
where ${g_\ell}(r)$ and ${h_\ell}(r)$ are given by 
\be
{g_\ell}(r) = {\int _0 ^r} {{\tilde r}^{\ell+2}} d {\tilde r} {\int _0 ^\pi} \sin \theta d \theta \phi({\tilde r},\theta){P_\ell}(\cos \theta)
\ee
\be
{h_\ell}(r) = {\int _r ^\infty} {{\tilde r}^{1-\ell}} d {\tilde r} {\int _0 ^\pi} \sin \theta d \theta \phi({\tilde r},\theta){P_\ell}(\cos \theta)
\ee

We will find it useful to examine the behavior of the curvature component $\varrho$, which for Brill wave initial data is given asymptotically (as $r \to \infty$) by the expression
\bea
{r^2} \varrho &=& - {r^2} {\frac {{\partial ^2}q} {\partial {r^2}}} - 
{\frac {{\partial ^2}q} {\partial {\theta^2}}} - \cot \theta {\frac {\partial q} {\partial \theta}} + 2 {r^2} {\frac {\partial q} {\partial r}} {\frac {\partial F} {\partial r}}
\nonumber
\\
&-& 2 {\frac {\partial q} {\partial \theta}} {\frac {\partial F} {\partial \theta}} - 4 {r^2} {\frac {{\partial ^2}F} {\partial {r^2}}} - 4 r {\frac {\partial F} {\partial r}} - 2 {\frac {{\partial ^2}F} {\partial {\theta^2}}}
\nonumber
\\
&-& 2 \cot \theta {\frac {\partial F} {\partial \theta}} - 4 {{\left ( {\frac {\partial F} {\partial \theta}} \right ) }^2}
\eea

We first choose $q$ to take the form
\be
q={a_0}{r^2}{\sin ^2}\theta {{({r^2}+{r_0^2})}^{-\gamma}}
\label{qformula}
\ee
where ${a_0},{r_0}$ and $\gamma$ are constants.  The rate of fall-off is determined by the constant $\gamma$. 

\begin{figure}
	\includegraphics[width=1.0\linewidth]{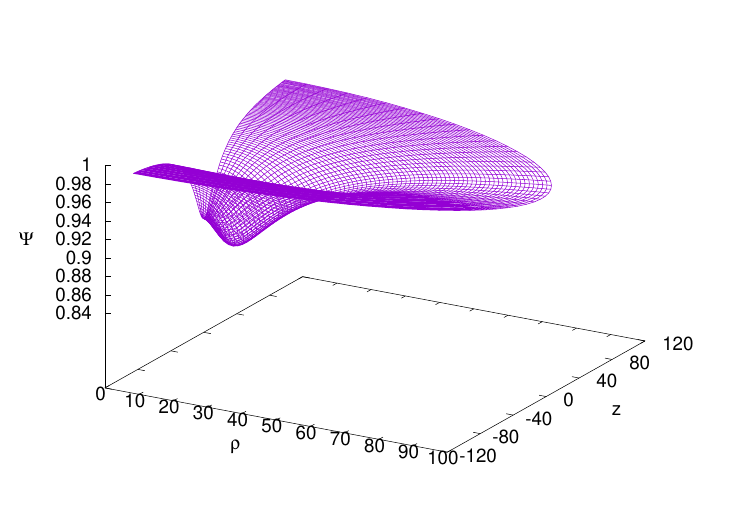}
	\caption{$\Psi$ with $q$ given by eqn. (\ref{qformula}) with ${a_0}=8, \, {r_0}=10, \, \gamma = 3/2$}			\label{psifig1}
\end{figure}

\begin{figure}
	\includegraphics[width=1.0\linewidth]{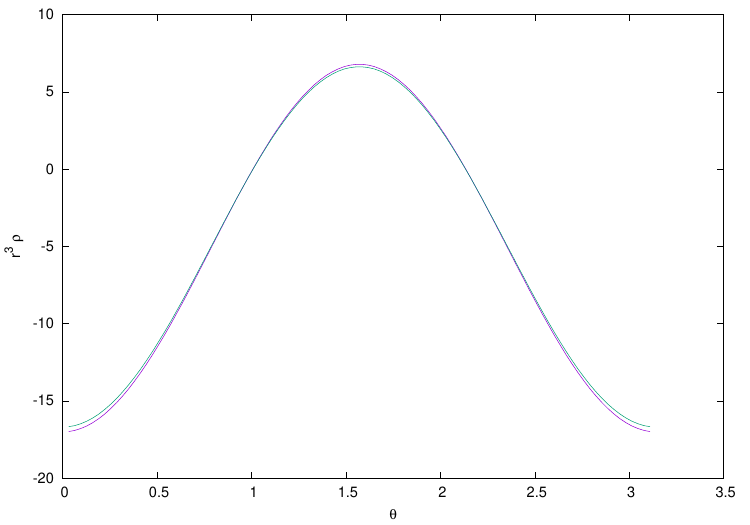}
	\caption{${r^3}\varrho$ as a function of $\theta$ for $r=10000$ and $r=15000$ with $q$ given by eqn. (\ref{qformula}) with ${a_0}=8, \, {r_0}=10, \, \gamma = 3/2$}			\label{cfig1}
\end{figure}

Figure (\ref{psifig1}) gives the numerically computed $\Psi$ for the case
${a_0}= 8, {r_0}=10, \gamma = 3/2$ up to the radius of $r=100$.  To examine the asymptotic behavior of $\varrho$ it is helfpul to go far out into the asymptotic region. Figure (\ref{cfig1}) plots ${r^3}\varrho$ for two different values of $r$: $r=10000$ and $r=15000$.   The fact that the two curves agree tells us that the asymptotic behavior of $\varrho$ is $\varrho \propto {r^{-3}}$

\begin{figure}
	\includegraphics[width=1.0\linewidth]{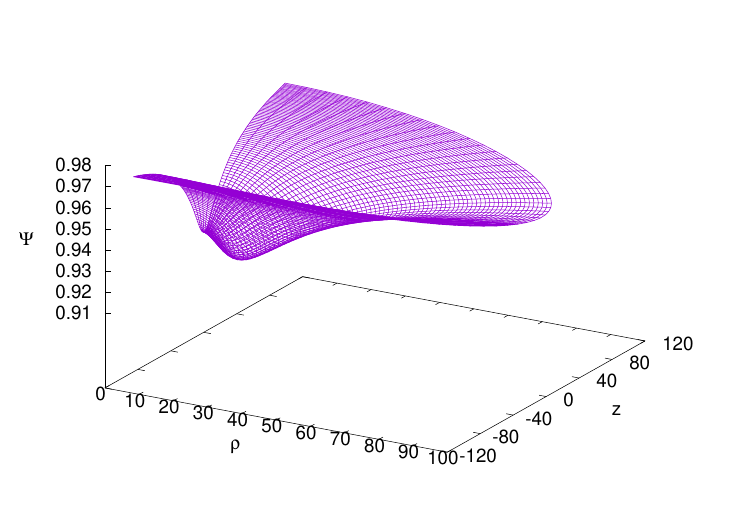}
	\caption{$\Psi$ with $q$ given by eqn. (\ref{qformula}) with ${a_0}=0.8, \, {r_0}=10, \, \gamma = 5/4$}			\label{psifig2}
\end{figure}

\begin{figure}
	\includegraphics[width=1.0\linewidth]{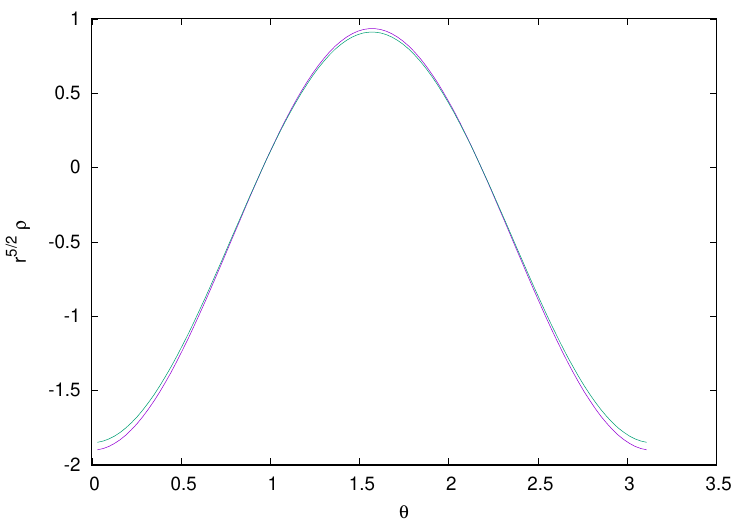}
	\caption{${r^{5/2}}\varrho$ as a function of $\theta$ for $r=10000$ and $r=15000$ with $q$ given by eqn. (\ref{qformula}) with ${a_0}=0.8, \, {r_0}=10, \, \gamma = 5/4$	}	\label{cfig2}
\end{figure}

Figure (\ref{psifig2}) gives the numerically computed $\Psi$ for the case
${a_0}= 0.8, {r_0}=10, \gamma = 5/4$ up to the radius of $r=100$.  Figure (\ref{cfig2}) plots ${r^{5/2}}\varrho$ for two different values of $r$: $r=10000$ and $r=15000$.   The fact that the two curves agree tells us that the asymptotic behavior of $\varrho$ is $\varrho \propto {r^{-5/2}}$
 
\begin{figure}
	\includegraphics[width=1.0\linewidth]{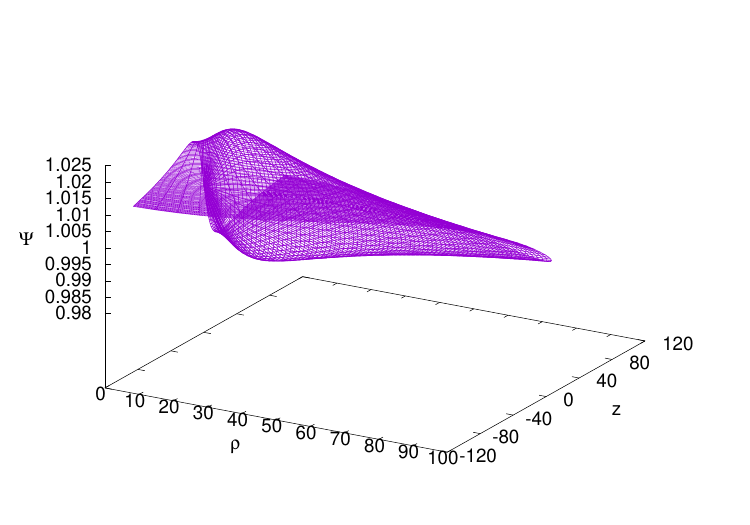}
	\caption{$\Psi$ with $q$ given by eqn. (\ref{qformula2}) with ${a_0}=1, \, {r_0}=10, \, \gamma = 7/4$}			\label{psifig3}
\end{figure}

\begin{figure}
	\includegraphics[width=1.0\linewidth]{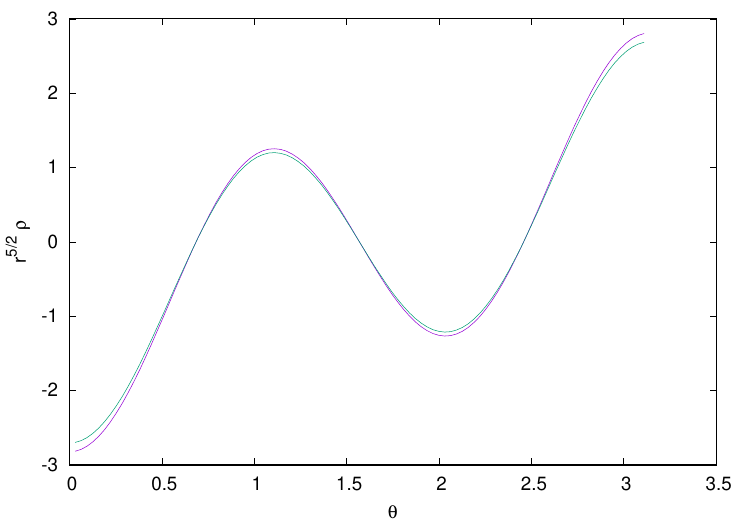}
	\caption{${r^{5/2}}\varrho$ as a function of $\theta$ for $r=10000$ and $r=15000$ with $q$ given by eqn. (\ref{qformula2}) with ${a_0}=1, \, {r_0}=10, \, \gamma = 7/4$	}	\label{cfig3}
\end{figure}

The choice of $q$ in eqn. (\ref{qformula}) is symmetric under $\theta \to \pi - \theta$, or to put it another way $q$ is symmetric under the antipodal transformation $(\theta,\phi) \to (\pi-\theta,2\pi-\phi)$ which takes each point on the two-sphere to its antipode.  In order to have an example that is not antipodally symmetric, we consider $q$ of the form
\be
q={a_0}{r^3}\cos \theta {\sin ^2}\theta {{({r^2}+{r_0^2})}^{-\gamma}}
\label{qformula2}
\ee
Figure (\ref{psifig3}) gives the numerically computed $\Psi$ for the case
${a_0}= 1, {r_0}=10, \gamma = 7/4$ up to the radius of $r=100$.  Figure (\ref{cfig3}) plots ${r^{5/2}}\varrho$ for two different values of $r$: $r=10000$ and $r=15000$.   The fact that the two curves agree tells us that the asymptotic behavior of $\varrho$ is $\varrho \propto {r^{-5/2}}$

Note from figure (\ref{cfig3}) that $\varrho$ is not antipodally symmetric.  This is relevant to the conjecture of Strominger\cite{strominger} that in the limit of early time at null infinity there is a symmetry under the combination of time reflection and antipodal mapping.  Since Brill wave initial data has zero extrinsic curvature, it follows that the time evolution of this initial data has time reflection symmetry.  Therefore the Brill waves plotted in figures (\ref{psifig3}-\ref{cfig3}) do not satisfy the conjecture of \cite{strominger}.  This result should not be regarded as a counterexample to this conjecture, but rather as a statement about the sort of asymptotic flatness under which this conjecture can hold: that is, the conjecture of \cite{strominger} does not hold for slow fall off.

\section{Conclusions}
\label{conclusions}

We have devised a simple method for generating Brill wave initial data of a given prescribed slow fall-off.  Our main result is a mathematical proof that the initial data do in fact have the particular slow fall-off that we prescribe.  In addition, we have presented a simple numerical method to solve for the conformal factor which completes the description of the initial data.

Since the usual Brill wave initial data has been a useful testbed for numerical evolution codes, it would be of interest to numerically evolve our slow fall-off Brill wave initial data, and in particular to see whether the slower fall-off in space and time at null infinity can be seen in the simulations.  One challenge to such a numerical evolution is that the usual outgoing wave boundary conditions used in numerical codes have been devised to be compatible with the usual (non-slow) fall-off of gravitational waves.  A related challenge has to do with the use of spectral methods: since the usual types of spectral basis functions are well adapted for the usual version of asymptotic flatness, it is not clear what spectral basis functions one should use for slow fall-off.

A byproduct of our study is an example of initial data of type (B) whose evolution cannot satisfy the antipodal conjecture of \cite{strominger}.  This example should not be regarded as a counterexample to the conjecture of \cite{strominger} but rather as an example of a limitation of the class of spacetimes to which the antipodal conjecture can apply. Since we have shown that type (A) data reduces to type (CK) under the time-symmetric Brill wave ansatz, it follows that antipodal symmetry holds for Brill wave data with stronger fall-off. A next task is to consider generalized data allowing non-zero extrinsic curvature, seeking an explicit example violating antipodal symmetry with stronger fall-off than achieved here. 
Indeed, since the (CK) spacetimes satisfy the antipodal conjecture, but only in a trivial sense, it would be interesting to delineate more precisely that class of spacetimes to which the antipodal conjecture applies in a nontrivial way. 
In an upcoming paper \cite{lz1}, it is shown that general spacetimes of type (A) need not have the antipodal symmetry property. However, boosted Schwarzschild and boosted Kerr spacetimes, or sums thereof, have this symmetry. 

\section*{Acknowledgements}
DG was supported by NSF grant PHY-2102914 to Oakland University.  LB was supported by NSF grant DMS-2204182 to the University of Michigan.  We would like to thank Andy Strominger and Manuela Campanelli for helpful discussions.

\appendix

\section{\small Curvature Notation}
\label{appinfo}

The curvature components are decomposed into components tangential to local $2$-spheres and null directions. 
Let $C_u$ denote an outgoing null hypersurface (that is an outgoing light cone), and $H_t$ a spacelike hypersurface of the spacetime manifold 
$(M, \mathbf{g})$. Then their intersection $S_{t.u} = H_t \cap C_u$ is diffeomorphic to the sphere. Let $e_A : A = \{ 1, 2 \}$ be an orthonormal frame of vectorfields tangential to $S_{t.u}$, let $\underline{L} = e_3$ be an incoming and $L = e_4$ an outgoing null vectorfield. These form 
a null frame $(e_1, e_2, e_3, e_4)$. For details see \cite{sta}, or \cite{lydia1}, \cite{lydia2}. \\

\begin{Def}
Let $\Pi$ denote the projection operator from the tangent space of $M$ to the tangent space of $S_{t,u}$. 
We define the null components of the Weyl curvature $W$ as follows: 
\bea
\underline{\alpha}_{\mu \nu}  \ & = & \ 
\Pi_{\mu}^{\ \rho} \ \Pi_{\nu}^{\ \sigma} \ W_{\rho \gamma \sigma \delta} \
e_3^{\gamma} \ e_3^{\delta} 
\label{underlinealpha} \\ 
\underline{\beta}_{\mu}  \ & = & \ 
\frac{1}{2} \ \Pi_{\mu}^{\ \rho} \ W_{\rho \sigma \gamma \delta} \  e_3^{\sigma} \
e_3^{\gamma} \ e_4^{\delta} 
\label{underlinebeta} \\ 
\varrho \ & = & \ 
\frac{1}{4} \ W_{\alpha \beta \gamma \delta} \ e_3^{\alpha} \ e_4^{\beta} \
e_3^{\gamma} \ e_4^{\delta} 
\label{rho} \\ 
\sigma  \ & = & \ 
\frac{1}{4} \ \ ^*W_{\alpha \beta \gamma \delta} \ e_3^{\alpha} \ e_4^{\beta} \
e_3^{\gamma} \ e_4^{\delta} 
\label{sigma} \\ 
\beta_{\mu}   \ & = & \  
\frac{1}{2} \ \Pi_{\mu}^{\ \rho} \ W_{\rho \sigma \gamma \delta} \ e_4^{\sigma} \
e_3^{\gamma} \ e_4^{\delta} 
\label{beta} \\ 
\alpha_{\mu \nu}  \ & = & \ 
\Pi_{\mu}^{\ \rho} \ \Pi_{\nu}^{\ \sigma} \ W_{\rho \gamma \sigma \delta} \
e_4^{\gamma} \ e_4^{\delta}  \ . 
\label{alphaR}
\eea
\end{Def}

Thus, capital indices taking the values $1,2$, we have: 
\bea
W_{A3B3} \ & = & \ \underline{\alpha}_{AB} \label{intnullcurvalphaunderline*1} \\ 
W_{A334} \ & = & \ 2 \ \underline{\beta}_A \\ 
W_{3434} \ & = & \ 4 \ \varrho \\ 
\ ^* W_{3434} \ & = & \ 4 \ \sigma \\ 
W_{A434} \ & = & \ 2 \ \beta_A \\ 
W_{A4B4} \ & = & \ \alpha_{AB}  \label{intnullcurvalpha*1}
\eea
with \\ 
\begin{tabular}{lll}
$\alpha$, $\underline{\alpha}$ & : & $S$-tangent, symmetric, traceless tensors \\ 
$\beta$, $\underline{\beta}$ & : &  $S$-tangent $1$-forms \\ 
$\varrho$, $\sigma$ & : & scalars \ . \\ 
\end{tabular}

\section{\small Weighted Sobolev Spaces}
\label{app:sobolev}

Here we briefly review the relevant definitions and essential properties of the weighted Sobolev spaces $W^{k,p}_\delta$ and $W^{\prime \, k,p}_\delta$, largely following the notation of Bartnik \cite{bartnik1}, to whom we refer the reader for further details. 

Given $p \in [1,\infty)$ and $\delta \in \mathbb R$ and setting $\sigma := \sqrt{1+r^2}$ with $r$ the radial coordinate on $\mathbb R^n$, we consider the norms
\bea
\|U\|_{p,\delta} & := & \left( \int_{\mathbb R^n} |U|^p \sigma^{-\delta p - n} d^n x \right)^{1/p},
\\
\|U\|^{\prime}_{p,\delta} & := & \left( \int_{\mathbb R^n} |U|^p r^{-\delta p - n} d^n x \right)^{1/p}
\eea
on functions $U \in L^p_{\text{loc}}(\mathbb R^n), \, L^p_{\text{loc}}(\mathbb R^n \backslash \{0\})$ respectively. We also allow for the limiting case $p = \infty$ by setting
\bea
\|U\|_{\infty,\delta} & := & \text{ess sup} \ \sigma^{-\delta} |U|,
\\
\|U\|^{\prime}_{\infty,\delta} & := & \text{ess sup} \ r^{-\delta} |U|.
\eea
The subspaces on which these norms are finite are the weighted Lebesgue spaces $L^p_\delta$, $L^{\prime \, p}_\delta$. Given $k \in \mathbb N_0$, the weighted Sobolev spaces $W^{k,p}_\delta$, $W^{\prime \, k,p}_\delta$ are the further subspaces on which the norms
\bea
\|U\|_{k,p,\delta} & := & \sum_{|\alpha| \leq k} \|D^\alpha U\|_{p,\delta-|\alpha|},
\\
\|U\|^{\prime}_{k,p,\delta} & := & \sum_{|\alpha| \leq k} \|D^\alpha U\|^{\prime}_{p,\delta-|\alpha|}
\eea
are finite, where $\alpha$ runs over multi-indices in the distributional derivatives $D^\alpha U$.

Among the most important properties of these function spaces, drawn on repeatedly in Section \ref{proof}, are the weighted H\"older and Sobolev inequalities (\cite{bartnik1}, Theorem 1.2):
\begin{prop} \label{prop:weighted}
\begin{enumerate}[(i)]
    \item If $U \in L^q_{\delta_1}$ and $V \in L^r_{\delta_2}$ with $1 \leq q,r \leq \infty$ and $\delta_1,\delta_2 \in \mathbb{R}$, then $UV \in L^p_\delta$ where $\delta := \delta_1 + \delta_2$ and $\frac{1}{p} := \frac{1}{q} + \frac{1}{r}$, with
    \be
    \|UV\|_{p,\delta} \leq \|U\|_{q,\delta_1} \cdot \|V\|_{r,\delta_2}.
    \ee
    \item Let $U \in W^{k,p}_\delta$. If $kp < n$, then $U \in L^{np/(n-kp)}_\delta$ with
    \be
    \|U\|_{np/(n-kp),\delta} < C \| U \|_{k,p,\delta}.
    \ee
    If $kp > n$, then $U \in L^\infty_\delta$ with
    \be
    \|U\|_{\infty,\delta} < C \| U \|_{k,p,\delta},
    \ee
    and in fact $U = o(r^{\delta})$ as $r \to \infty$.
\end{enumerate}
\end{prop} 
Note that, by taking $V = 1$, the first of these implies that  $L^q_{\delta_2} \subset L^p_{\delta_1}$ for $1 \leq p \leq q \leq \infty$ and $\delta_1 > \delta_2$.

Our results hinge upon the simplicity of the Laplacian operator $\Delta$, together with the fact that our operators of interest are ``close" to $\Delta$ in the following sense (\cite{bartnik1}, Definition 1.5):

\begin{Def} \label{def:asymptotic}
    An operator $U \mapsto PU$ defined by
    \be
    PU := a^{ij}(x) \partial^2_{ij} U + b^i(x) \partial_i U + c(x) U
    \ee
    is said to be \underline{asymptotic to $\Delta$} at rate $\tau > 0$ and exponent $2p$, $n < 2p < \infty$, if there exists a $\lambda > 0$ such that
    \be
    \lambda |\xi|^2 \leq a_{ij}(x)\xi^i \xi^j \leq \lambda^{-1} |\xi|^2
    \ee
    for all $x, \xi \in \mathbb R^n$ and the coefficient functions satisfy
    \bea
    (a_{ij} - \delta_{ij}) \in & W^{1,2p}_{-\tau},
    \\ \ b^i \in & L^{2p}_{-1-\tau},
    \\ \ c \in & L^{p}_{-2-\tau}.
    \eea
\end{Def}
In particular, this definition ensures that $P$ is bounded as an operator $W^{2,s}_\delta \to W^{0,s}_{\delta-2}$ for any $1 \leq s \leq p$ and $\delta \in \mathbb R$.

\end{document}